\documentclass[sort&compress
    ,final            % use final for the camera ready runs
  ]
  {aipproc}

\layoutstyle{8x11single}
   \usepackage{showkeys}
    \usepackage{amssymb}
    \usepackage{bm}
    \usepackage{amsmath}

\newcommand\beq{\begin{equation}}
\newcommand\eeq{\end{equation}}
\newcommand\beqa{\begin{eqnarray}}
\newcommand\eeqa{\end{eqnarray}}

\newcommand{\dd}{\text{d}}
\newcommand{\ee}{\text{e}}
\newcommand{\hcs}{\text{hcs}}

\begin{document}

\title{A simple model kinetic equation for inelastic Maxwell particles}


\author{Andr\'es Santos}{
  address={Departamento de F\'{\i}sica, Universidad de Extremadura, Badajoz,
  Spain}
}

%\author{<author2>}{  address={<common address for author2 and author3>}}

%\author{<author3>}{  address={<common address for author2 and author3>} ,altaddress={<author1 address>}
% additional visiting address}

\begin{abstract}
The model of inelastic Maxwell particles (IMP)  allows one to derive
some exact results which show  the strong influence of inelasticity
on the nonequilibrium properties of a granular gas. The aim of this
work is to propose a simple model kinetic equation that preserves
the most relevant properties of the Boltzmann equation (BE) for IMP
and reduces to the  BGK  kinetic model in the elastic limit. In the
proposed kinetic model the collision operator is replaced by a
relaxation-time term toward a reference Maxwellian distribution plus
a term representing the action of a  friction force. It contains
three parameters (the relaxation rate, the effective temperature of
the reference Maxwellian, and the friction coefficient) which are
determined by imposing consistency with basic exact properties of
the BE for IMP. As a consequence, the kinetic model reproduces the
true shear viscosity and predicts  accurate expressions for the
transport coefficients associated with the heat flux.  The model can
be exactly solved for the homogeneous cooling state, the solution
exhibiting an algebraic high-energy tail with an exponent in fair
agreement with the correct one.

\end{abstract}

\maketitle

%%%%%%%%%%%%%%%%%%%%%%%%%%%%%%%%%%%%%%%%%%%%
%% MAINMATTER
%%%%%%%%%%%%%%%%%%%%%%%%%%%%%%%%%%%%%%%%%%%%

\section{Introduction}
The prototype model for the description of granular media in the
regime of rapid flow consists of an assembly of (smooth) inelastic
hard spheres (IHS) with a constant coefficient of normal restitution
$\alpha<1$. In the low density limit the velocity distribution
function obeys the Boltzmann equation (BE), modified to account for
the inelasticity of collisions \cite{GS95}. Because of the
mathematical intricacy of the BE for IHS, a simpler model of
inelastic Maxwell particles (IMP)  has been proposed
\cite{BCG00,BNK00,KBN02}, where the collision rate is assumed to be
independent of the relative speed of the colliding pair. Apart from
its interest as a model of granular gases, the IMP model is
interesting by itself since it allows the derivation of some
\emph{exact} results for any dimensionality $d$, showing
unambiguously the strong influence of inelasticity  on the
nonequilibrium properties of the gas.

While the BE for IMP is considerably simpler than for IHS, it is
still a formidable task to solve it in a closed form, even in the
case of the homogeneous cooling state (HCS). The aim of this work is
to propose a simple model kinetic equation that preserves the most
relevant properties of the BE for IMP and reduces to the celebrated
Bhatnagar--Gross--Krook (BGK) model in the elastic limit $\alpha\to
1$. In the proposed kinetic model the collision operator is replaced
by a relaxation-time term toward a reference Maxwellian distribution
plus a term representing the action of a dissipative friction force.
The kinetic model contains three parameters:  a relaxation rate
modified by a factor $\beta(\alpha)$ with respect to its elastic
value,  an effective reference temperature modified by a factor
$\theta(\alpha)$ with respect to the actual granular temperature,
and  a friction coefficient $\gamma(\alpha)$. The model is a hybrid
of two previous models \cite{BMD96,BDS99} originally proposed for
IHS, reducing to them if either $\gamma(\alpha)=0$ or
$\theta(\alpha)=1$, respectively. The three parameters
$\beta(\alpha)$, $\theta(\alpha)$, and $\gamma(\alpha)$ are
determined by imposing consistency with  basic exact properties
derived from the BE for IMP. As a consequence, the model reproduces
the true shear viscosity and predicts  accurate expressions for the
transport coefficients associated with the heat flux. Moreover, it
can be exactly solved for the HCS, the solution exhibiting an
algebraic high-energy tail with an exponent
 in fair agreement with the correct one, especially for high
inelasticity.

\section{The Boltzmann equation for inelastic Maxwell particles}
The BE for IMP \cite{BCG00,BNK00,KBN02} can be obtained from the BE
for IHS by replacing the term $|\mathbf{g}\cdot
\widehat{\bm{\sigma}}|$ in the collision rate (where ${\bf g}={\bf
v}_1-{\bf v}_2$ is the relative velocity  and
$\widehat{\bm{\sigma}}$ is the unit vector directed along the
centers of the two colliding spheres) by an {\em average\/} value
proportional to the thermal speed $\sqrt{2T/m}$ (where $T$ is the
granular temperature and $m$ is the mass of a particle). In the
version of the IMP model first proposed by Bobylev \emph{et al.}
\cite{BCG00} the collision rate has the same dependence on the
scalar product $\widehat{\mathbf{g}}\cdot \widehat{\bm{\sigma}}$ as
in the case of hard spheres. In a simpler version
\cite{BNK00,KBN02},  the collision rate is assumed to be independent
of $\widehat{\mathbf{g}}\cdot \widehat{\bm{\sigma}}$. In this latter
case, the corresponding BE reads
\beq
(\partial_t+{\bf v}_1\cdot\nabla)f({\bf r},{\bf v}_1;t)=J[{\bf
r},{\bf v}_1;t|f]\equiv\frac{d+2}{2}\frac{\nu_0({\bf r},t)}{n({\bf
r},t)\Omega_d}\int \dd \widehat{\bm{\sigma}}\int \dd {\bf v}_2
\left(\alpha^{-1}\mathfrak{b}^{-1}-1\right) f({\bf r},{\bf
v}_1;t)f({\bf r},{\bf v}_2;t) ,
\label{1}
\eeq
where  $n$ is the number density, $\nu_0\propto nT^{1/2}$
 is an effective collision
frequency whose specific form will not be needed,
$\Omega_d=2\pi^{d/2}/\Gamma(d/2)$ is the total solid angle in $d$
dimensions,  and $\mathfrak{b}$ is the operator transforming
pre-collision velocities into post-collision ones, namely,
$\mathfrak{b}{\bf v}_{1,2}={\bf v}_{1,2}\mp{(1+\alpha)}({\bf
g}\cdot\widehat{\bm{\sigma}})\widehat{\bm{\sigma}}/2$. The Boltzmann
collision operator $J[f]$ conserves mass and momentum but not
energy. The collisional moments of second and third degree are
\cite{S03}
\beq
\frac{m}{d}\int \dd\mathbf{v}\,
 V^2 J[f]=-\zeta p,\quad  m\int
\dd\mathbf{v}\,
\left(V_iV_j-{d}^{-1}V^2\delta_{ij}\right)J[f]=-\nu_\eta
\left(P_{ij}-p\delta_{ij}\right),\quad \frac{m}{2}\int
\dd\mathbf{v}\, V^2\mathbf{V}J[f]=-\nu_\kappa\mathbf{q}.
\label{3}
\eeq
Here, $\mathbf{V}\equiv \mathbf{v}-\mathbf{u}$ is the peculiar
velocity, where $\mathbf{u}$ is the flow velocity, $ \mathsf{P}$ is
the pressure tensor,  $p=nT=d^{-1}\text{tr}\, \mathsf{P}$ is the
hydrostatic pressure, and $\mathbf{q}$ is the heat flux. The exact
expressions for the cooling rate $\zeta$ and the rates of change
$\nu_\eta$ and $\nu_\kappa$ are \cite{S03}
\beq
\zeta^*\equiv \zeta/\nu_0=
\frac{d+2}{4d}\left(1-\alpha^2\right),\quad \nu_\eta^*\equiv
\nu_\eta/\nu_0=\frac{\left(1+\alpha\right)^2}{4}+\zeta^*,\quad
\nu_\kappa^*\equiv
\nu_\kappa/\nu_0=\frac{\left(1+\alpha\right)^2(d-1)}{4d}+\frac{3}{2}\zeta^*.
\label{6}
\eeq
 In isotropic states, the collisional
moment of fourth degree is
\beq
n^{-1}\int\dd \mathbf{v} \, V^4 J[f]=-\nu_2 M_2+\lambda
\nu_0(2T/m)^2, \quad M_\ell\equiv n^{-1} \int\dd \mathbf{v} \,
V^{2\ell} f,
\label{8}
\eeq
where the exact expressions for the rate of change $\nu_2$ and the
coefficient $\lambda$ are \cite{S03}
\beq
\nu_2^*\equiv \nu_2/\nu_0
={\left(1+\alpha\right)^2(4d-7+6\alpha-3\alpha^2)}/{16d}+2\zeta^*,
\quad\lambda={\left(1+\alpha\right)^2(d+2)(4d-1-6\alpha+3\alpha^2)}/{64}.
\label{10}
\eeq

In the case of the uniform, free cooling state, the BE (\ref{1}) and
the evolution equations for the second- and fourth-degree moments
become
\beq
\partial_t f(\mathbf{v})=J[\mathbf{v}|f],\quad
\partial_t T=-\zeta T,\quad \partial_t M_2=-\nu_2 M_2+\lambda\nu_0
(2T/m)^2.
\label{12}
\eeq
The solution to the cooling equation is
$T(t)=T(0)[1+\zeta(0)t/2]^{-2}$ (Haff's law). If time is measured by
the accumulated number of collisions per particle as
$\tau(t)=\int_0^t\dd t'\, \nu_0(t')=(2/\zeta^*)\ln[1+\zeta(0)t/2]$,
Haff's law becomes $T(\tau)=T(0)\exp(-\zeta^*\tau)$. It is
convenient to introduce the \emph{reduced} moments $M_\ell^*\equiv
M_\ell /(2T/m)^\ell$, so that $M_1^*=d/2$ by definition, and the
\emph{reduced} distribution $f^*(\mathbf{c},\tau)$ defined by
\beq
f(\mathbf{v},t)=n[m/2T(t)]^{d/2} f^*(\mathbf{c},\tau),\quad
\mathbf{c}=\mathbf{v}/\sqrt{2T(t)/m}.
\label{15}
\eeq
Thus, Eq.\ \eqref{12} yields
\beq
\partial_\tau f^*(\mathbf{c})+({\zeta^*}/{2})\partial_\mathbf{c}\cdot\mathbf{c}
f^*(\mathbf{c})=J^*[\mathbf{c}|f^*], \quad
\partial_\tau M_2^*=-(\nu_2^*-2\zeta^*)M_2^*+\lambda,
\label{14}
\eeq
where $J^*[\mathbf{c}|f^*]$ is the dimensionless version of the
collision operator $J[\mathbf{v}|f]$. Except in the one-dimensional
case, the reduced moment $M_2^*(\tau)$ converges in time to the well
defined value $M_2^*(\infty)=\lambda/(\nu_2^*-2\zeta^*)$. In
general, the distribution function reaches a \emph{scaling form}
\cite{BC03}, called homogeneous cooling state (HCS), which  is  the
stationary solution of Eq.\ \eqref{14}, i.e.,
$f^*(\mathbf{c},\tau)\to
f^*(\mathbf{c},\infty)=f^*_\hcs(\mathbf{c})$. Its exact expression
is not known, except in the one-dimensional case, where
$f^*_\hcs(\mathbf{c})=(2^{3/2}/\pi)(1+2c^2)^{-2}$ \cite{BMP01}. For
$d\geq 2$, the fourth cumulant (or kurtosis) $a_2\equiv 4M_2^*
/d(d+2)-1$ of the distribution $f^*_\hcs$ is
\beq
a_2=\frac{4}{d(d+2)}\frac{\lambda}{\nu_2^*-2\zeta^*}-1=\frac{6(1-\alpha)^2}{4d-7+6\alpha-3\alpha^2}.
\label{16}
\eeq
Therefore, $a_2\geq 0$ for IMP, in contrast to what happens in the
case of IHS with $\alpha> \sqrt{2}/2$ \cite{vNE98}. It is also known
that $f^*_\hcs$ exhibits an \emph{algebraic} high-energy tail of the
form \cite{KBN02}
\beq
f^*_\hcs(\mathbf{c})\sim c^{-d-s(\alpha)},
\label{18}
\eeq
where the exponent $s(\alpha)$ is the solution of the transcendental
equation
\beq
1-\frac{1-\alpha^2}{4d}
s=_2\!\!F_1[-s/2,1/2;d/2;1-(1-\alpha)^2/4]+\left(\frac{1+\alpha}{2}\right)^s\frac{\Gamma(s/2+1/2)\Gamma(d/2)}{\Gamma(s/2+d/2)\Gamma(1/2)},
\label{19}
\eeq
$_2F_1[a,b;c;z]$ being a hypergeometric function. Equation
\eqref{18} implies that those moments $M_\ell^*$ with $\ell\geq
s(\alpha)/2$ are divergent.

%\beq
%P_{ij}=p\delta_{ij}-\eta\left(\nabla_i u_j+\nabla_j
%u_i-2d^{-1}\nabla\cdot\mathbf{u}\delta_{ij}\right),\quad
%\mathbf{q}=-\kappa \nabla T-\mu\nabla n,
%\label{20}
%\eeq
Finally,  the exact expressions for the transport coefficients
 are \cite{S03}
\beq
\eta=\frac{p}{\nu_0}\frac{1}{\nu_\eta^*-\zeta^*/2},\quad
\kappa=\frac{d+2}{2}\frac{p}{m\nu_0}\frac{1+2a_2}{\nu_\kappa^*-2\zeta^*},\quad
\mu=\frac{2T}{n}(\kappa-\kappa'),\quad
\kappa'=\frac{d+2}{2}\frac{p}{m\nu_0}\frac{1+3a_2/2}{\nu_\kappa^*-3\zeta^*/2}.
\label{21}
\eeq
Here, $\eta$ is the shear viscosity, $\kappa$ is the thermal
conductivity, $\kappa'$ is a modified thermal conductivity, and
$\mu$ is a coefficient relating $\mathbf{q}$ and $\nabla n$. As can
be seen from \eqref{6}, $\nu_\eta^*>\zeta^*/2$ and
$\nu_\kappa^*>3\zeta^*/2$, so that  $\eta$ and  $\kappa'$ are well
defined for all $\alpha$. On the other hand,  $\kappa$ and  $\mu$
are not positive definite if $\nu_\kappa^*\leq 2\zeta^*$, i.e., if
$\alpha\leq \alpha_h\equiv (4-d)/3d$. Therefore, if $d<4$ and
$\alpha\leq \alpha_h$, there is no hydrodynamic behavior since the
heat flux does not relax to a Fourier-law form.

\section{Model kinetic equation}
Although the BE for IMP is more manageable than for IHS and some
important properties are accessible in an exact way, its explicit
solution is not known, even for the HCS.  It is then natural to
wonder whether a simple generalization of the well-known BGK model
kinetic equation to the case of IMP can be proposed. The model
considered in this paper is constructed by the replacement
\beq
J[\mathbf{r},\mathbf{v};t|f]\to
\widetilde{J}[\mathbf{r},\mathbf{v};t|f]\equiv
-\beta(\alpha)\nu_0(\mathbf{r},t)\left[f(\mathbf{r},\mathbf{v};t)-f_0(\mathbf{r},\mathbf{v};t)\right]+{\gamma(\alpha)}
\nu_0(\mathbf{r},t)\partial_\mathbf{v}\cdot
\mathbf{V}f(\mathbf{r},\mathbf{v};t),
\label{3.1}
\eeq
where
\beq
f_0(\mathbf{r},\mathbf{v};t)=n(\mathbf{r},t)\left[m/2\pi\theta(\alpha)
T(\mathbf{r},t)\right]^{d/2}\exp\left[-mV^2/2\theta(\alpha)
T(\mathbf{r},t)\right]
\label{3.2}
\eeq
is a local equilibrium distribution parameterized by the temperature
$\theta(\alpha) T$.  The effect of the inelastic collisions in the
original BE is played in the model (\ref{3.1}) by  a relaxation term
toward the distribution $f_0$ at an effective temperature $\theta
T$, plus an external friction term. The model contains three free
parameters: the factor $\beta(\alpha)>0$ modifying the relaxation
rate with respect to its elastic value, the factor
$\theta(\alpha)<1$ modifying the granular temperature in the
reference state $f_0$, and the friction coefficient
$\gamma(\alpha)>0$. These three parameters will be chosen in the
next section to optimize the agreement  with the most important
properties of the BE for IMP. If one particularizes to
$\gamma(\alpha)=0$, the model (\ref{3.1}) reduces to the one
proposed by Brey, Moreno, and Dufty (BMD) \cite{BMD96}, while the
choice $\theta(\alpha)=1$ yields a simplified version of the model
proposed by Brey, Dufty, and Santos (BDS) \cite{BDS99}. {}From that
point of view, the model \eqref{3.1} can be seen as a hybrid of the
BMD and BDS models. Although the two latter models were originally
proposed for IHS, Eq.\ \eqref{3.1} is proposed here as a model for
IMP, not IHS. In fact, it has been shown \cite{BE04,DBZ04} that the
BMD model shares more features with the BE for IMP than for IHS.

Let us now obtain the basic physical properties of the model
operator $\widetilde{J}$. First, it is straightforward to check that
it conserves mass and momentum. However, since $\theta\neq 1$ and
$\gamma\neq 0$, energy is not conserved by collisions. More
specifically, the cooling rate  and the rates of change defined by
Eq.\ \eqref{3} are given (in reduced units) by
\beq
\widetilde{\zeta}^*(\alpha)=\beta(\alpha)[1-\theta(\alpha)]+2\gamma(\alpha),
\quad
\widetilde{\nu}_\eta^*(\alpha)=\beta(\alpha)\theta(\alpha)+\widetilde{\zeta}^*(\alpha),\quad
\widetilde{\nu}_\kappa^*(\alpha)=\frac{\beta(\alpha)}{2}\left[3\theta(\alpha)-1\right]+\frac{3}{2}\widetilde{\zeta}^*(\alpha),
\label{3.3}
\eeq
where henceforth a tilde means that the corresponding quantity has
been evaluated with the model operator $\widetilde{J}$. Moreover,
\beq
n^{-1}\nu_0^{-1}\int \dd
\mathbf{v}V^{2\ell}\widetilde{J}[f]=-\left(\beta+2\ell\gamma\right)M_\ell+\beta\frac{\Gamma(\ell+d/2)}{\Gamma(d/2)}
\left(\frac{2\theta T}{m}\right)^\ell.
\label{3.5}
\eeq
In particular, setting $\ell=2$ we reobtain Eq.\ \eqref{8} with
\beq
\widetilde{\nu}_2^*(\alpha)=\beta(\alpha)[2\theta(\alpha)-1]+2\widetilde{\zeta}^*(\alpha),\quad
\widetilde{\lambda}(\alpha)=\beta(\alpha)\theta^2(\alpha) d(d+2)/4.
\label{3.6}
\eeq

Let us now consider the free cooling problem. As shown by Eq.\
\eqref{14}, the necessary and sufficient condition to reach a
scaling solution (HCS) with a finite fourth-degree velocity moment
is $\widetilde{\nu}_2^*>2\widetilde{\zeta}^*$, i.e., $\theta>1/2$.
In that case, the first equality of Eq.\ \eqref{16} implies that the
fourth cumulant of the scaling solution is
\beq
\widetilde{a}_2(\alpha)={[1-\theta(\alpha)]^2}/\left[{2\theta(\alpha)-1}\right].
\label{3.7}
\eeq
Therefore, we have $\widetilde{a}_2>0$, regardless of the precise
values of the parameters $\beta$, $\theta$, and $\gamma$ of the
model. As a consequence, the model \eqref{3.1} can never capture the
negative values  exhibited by $a_2$ in the case of IHS for
$\alpha>\sqrt{2}/2$. {}From Eq.\ \eqref{3.5} it is easy to obtain
the time-dependence of the reduced velocity moments in the free
cooling problem:
\beq
M_\ell^*(\tau)=\ee^{-\beta\left[1-\ell(1-\theta)\right]\tau}M_\ell^*(0)+\left\{1-
\ee^{-\beta\left[1-\ell(1-\theta)\right]\tau}
\right\}\frac{\theta^\ell}{1-\ell(1-\theta)}\frac{\Gamma(\ell+d/2)}{\Gamma(d/2)}.
\label{3.8}
\eeq
Since, on physical grounds,  $\theta<1$, it turns out that the
reduced moments $M_\ell^*(\tau)$ \emph{diverge} in time if $\ell\geq
(1-\theta)^{-1}$. This implies that the scaling solution presents a
high-energy tail of the form \eqref{18} with the exponent
\beq
\widetilde{s}(\alpha)=2[1-\theta(\alpha)]^{-1}.
\label{3.9}
\eeq
Finally, the transport coefficients of the model are given by
\eqref{21}, except for the obvious replacements $\zeta^*\to
\widetilde{\zeta}^*$, $\nu_\eta^*\to \widetilde{\nu}_\eta^*$,
$\nu_\kappa^*\to \widetilde{\nu}_\kappa^*$,
 and $a_2\to\widetilde{a}_2$.

The main advantage of a kinetic model lies in the possibility of
obtaining the explicit form of the velocity distribution function.
Let us illustrate this in  the free cooling case. According to the
model (\ref{3.1}), the first equation of \eqref{14} becomes
\beq
\left\{\beta^{-1}\partial_\tau
+1+\left[({1-\theta})/{2}\right]{\partial_\mathbf{c}}\cdot\mathbf{c}
\right\}f^*(\mathbf{c})=(\pi\theta)^{-d/2} \exp(-c^2/\theta).
\label{3.10}
\eeq
 It is interesting to note that the parameter $\gamma$ does not
intervene explicitly in Eq.\ \eqref{3.10}, so that it is formally
equivalent to the equation obtained from the BMD model \cite{BMD96}.
Given an arbitrary initial condition $f^*(\mathbf{c},0)$, the exact
solution to Eq.\ \eqref{3.10} is \cite{DBZ04,BD06}
\beq
f^*(\mathbf{c},\tau)=\ee^{-\beta[1+d(1-\theta)/2]\tau}f^*\left(\ee^{-\beta(1-\theta)\tau/2}\mathbf{c},0\right)+
(\pi\theta)^{-d/2}\int_0^{\beta\tau}\dd y\,
\exp\left\{-[1+d(1-\theta)/2]y-c^2e^{-(1-\theta)y}/\theta\right\}.
\label{3.11}
\eeq
The HCS is obtained taking the limit $\tau\to\infty$ with the result
\cite{BMD96,DBZ04,BD06}
\beq
f^*_\hcs(\mathbf{c})=(\pi\theta)^{-d/2}(1-\theta)^{-1}(\theta/c^2)^{d/2+1/(1-\theta)}\left[\Gamma\left(d/2+1/(1-\theta)\right)
-\Gamma\left(d/2+1/(1-\theta),c^2/\theta\right)\right],
\label{3.12}
\eeq
where the change of variable $x=c^2\exp[-(1-\theta)y]/\theta$ has
been made and $\Gamma(z,a)=\int_a^\infty \dd x \,x^{z-1}\ee^{-x}$ is
the incomplete gamma function. Note that the whole dependence of
$f^*_\hcs$ on inelasticity appears through the parameter $\theta$
only. In the high-energy limit (actually, if $c^2\gg \theta$),
$\Gamma\left(d/2+1/(1-\theta),c^2/\theta\right)\to 0$, so that we
recover the tail \eqref{18} with the exponent \eqref{3.9}. The
transient from the initial distribution $f^*(\mathbf{c},0)$ to the
asymptotic distribution $f^*_\hcs(\mathbf{c})$ is described by Eq.\
\eqref{3.11}, which can be rewritten in a simpler form by
introducing the deviation $\delta
f^*(\mathbf{c},\tau)=f^*(\mathbf{c},\tau)-f^*_\hcs(\mathbf{c})$. The
results for $\delta f^*$ and its moments $\delta M_\ell^*$ with
$\ell<(1-\theta)^{-1}$ are
\beq
\delta f^*(\mathbf{c},\tau)=\ee^{-\beta[1+d(1-\theta)/2]\tau}\delta
f^*\left(\ee^{-\beta(1-\theta)\tau/2}\mathbf{c},0\right),\quad
\delta M_\ell^*(\tau)=
\ee^{-\beta\left[1-\ell(1-\theta)\right]\tau}\delta M_\ell^*(0).
\label{3.13}
\eeq
 While $f^*(\mathbf{c},\tau)$ relaxes uniformly  to $f^*_\hcs
(\mathbf{c})$ at fixed $\mathbf{c}$ with a relaxation rate
$\beta[1+d(1-\theta)/2]$, the moment $M_\ell^*(\tau)$ relaxes to its
HCS value with a  shorter relaxation rate
$\beta\left[1-\ell(1-\theta)\right]$ which goes to zero as $\ell$
approaches the threshold value $(1-\theta)^{-1}$ from below. This
paradoxical property \cite{BD06} is a consequence of the
non-commutability of the limits $c\to\infty$ and $\tau\to\infty$ in
Eq.\ \eqref{3.13}. To analyze this with more detail, let us suppose
that the initial distribution $f^*(\mathbf{c},0)$ has a high-energy
tail much weaker than that of $f_\hcs^*(\mathbf{c})$, so that
$\delta f^*(\mathbf{c},0)\approx -f_\hcs^*(\mathbf{c})\approx -A
c^{-d-2/(1-\theta)}$ in the region $c^2\gg \theta$, where the
amplitude $A$ is known from Eq.\ \eqref{3.12}. In that case, at
fixed $\tau$, one has
\beq
\delta f^*(\mathbf{c},\tau)\approx -A c^{-d-2/(1-\theta)}\approx -
f_\hcs^*(\mathbf{c}),\quad c^2\gg \theta e^{\beta(1-\theta)\tau}.
\label{3.15}
\eeq
This means that, at any fixed time $\tau$, there always exists an
infinite range of large speeds, $c^2\gg \theta
e^{\beta(1-\theta)\tau}$, where the deviation of the velocity
distribution function from its asymptotic HCS value is of 100\%.
This region is eventually responsible for the divergence of moments
with $\ell\geq (1-\theta)^{-1}$.

\section{Results and comparison with the Boltzmann
equation}

So far, all the properties of the kinetic model \eqref{3.1}
described in the preceding section are valid with independence of
the specific expressions for the parameters $\beta(\alpha)$,
$\theta(\alpha)$, and $\gamma(\alpha)$. Now we fix them by requiring
the model to reproduce the basic properties of the original BE. The
most characteristic consequence of inelasticity is the cooling rate,
so that an obvious requirement is $\zeta^*=\widetilde{\zeta}^*$. The
next requirement could be the agreement with either the relaxation
rate $\nu_\eta^*$ or $\nu_\kappa^*$. We cannot enforce both since
that would imply $\theta=d/(d+2)$, which yields an unphysical model
in the elastic limit. This impossibility of satisfying the shear
viscosity and the thermal conductivity simultaneously, which also
happens with the conventional BGK model, is remedied by a more
sophisticated model \cite{DBZ04}.  In the case of the kinetic model
\eqref{3.1}, let us take $\nu_\eta^*=\widetilde{\nu}_\eta^*$ as a
second condition. Finally, given the important role played by the
kurtosis $a_2$ of the HCS, the third condition adopted here is
$a_2=\widetilde{a}_2$. These three requirements yield
\beq
\beta(\alpha)=\frac{(1+\alpha)^2}{4\theta(\alpha)},\quad
\gamma(\alpha)=\frac{d+2}{8d}(1-\alpha^2)-\frac{(1+\alpha)^2}{8}\left[\theta^{-1}(\alpha)-1\right],\quad
\theta(\alpha)=1+a_2(\alpha)-\sqrt{a_2(\alpha)\left[1+a_2(\alpha)\right]},
\label{4.1}
\eeq
where $a_2(\alpha)$ is given by Eq.\ \eqref{16}.

\begin{figure}
\includegraphics[width=\columnwidth]{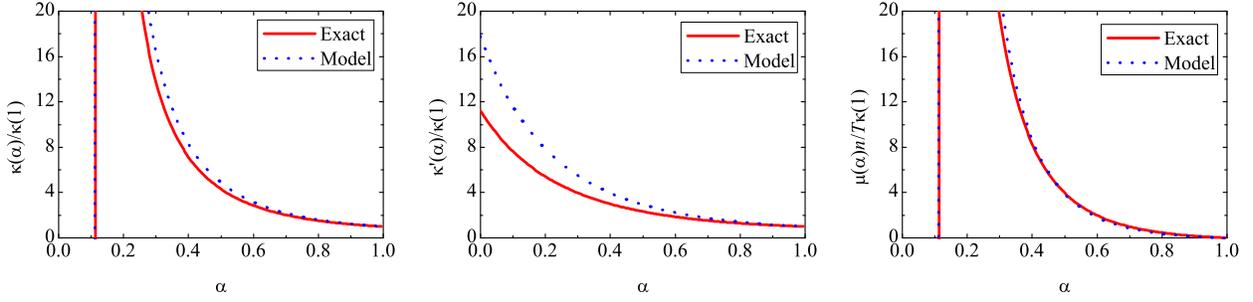}
\caption{Transport coefficients $\kappa(\alpha)$, $\kappa'(\alpha)$,
and $\mu(\alpha)$ for $d=3$.
\label{transport}}
\end{figure}
By construction, the model reproduces the exact shear viscosity
$\eta$ but not the  transport coefficients associated with the heat
flux. However, as Fig.\ \ref{transport} shows, the model captures
reasonably well the rapid increase of $\kappa(\alpha)$,
$\kappa'(\alpha)$, and $\mu(\alpha)$ with the inelasticity, the
agreement being especially remarkable in the case of $\mu(\alpha)$.
The model also predicts the existence of a threshold value
$\alpha_h$ below which hydrodynamics  no longer holds. The value of
$\alpha_h$ is the solution of a quartic equation which for $d=3$
yields $\alpha_h\simeq 0.114$, in excellent agreement with the exact
result $\alpha_h=1/9\simeq 0.111$.

\begin{figure}
\includegraphics[width=\columnwidth]{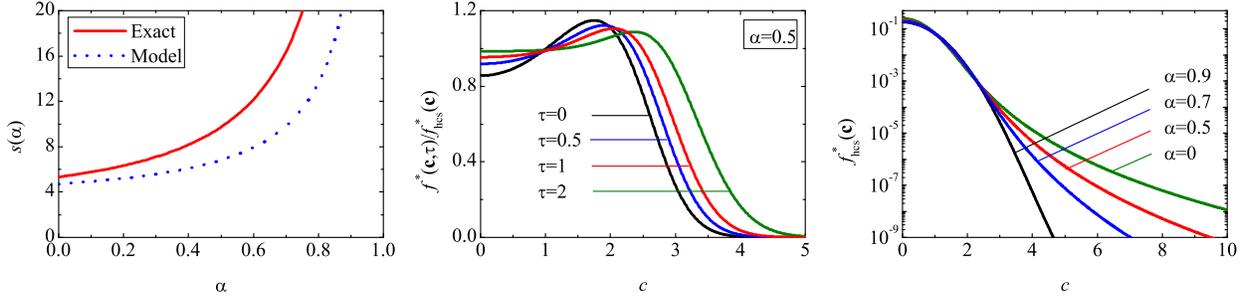}
\caption{Left panel: High-energy exponent $s(\alpha)$  for $d=3$.
Middle panel: Time evolution of the ratio
$f^*(\mathbf{c},\tau)/f^*_\hcs(\mathbf{c})$, starting from an
initial equilibrium distribution, for $\alpha=0.5$ and $d=3$. Right
panel: Plot of $f^*_\hcs(\mathbf{c})$ for $d=3$ and $\alpha=0$,
$0.5$, $0.7$, and $0.9$.
\label{evol}}
\end{figure}
As Fig.\ \ref{evol} shows, the model underestimates the exact
exponent $s(\alpha)$ but the agreement improves as the inelasticity
increases. The time evolution of the distribution function in the
free cooling state, as well as the asymptotic scaling solution are
also shown in Fig.\ \ref{evol} for a few representative cases.

\section{Concluding remarks}
The model of IMP shares with the more realistic model of IHS the
description of the important influence of inelasticity on the
dynamical properties of a granular gas. However, that influence is
magnified by the IMP model, giving rise to stronger departures from
the Maxwellian distribution and even to the absence of hydrodynamics
for sufficiently inelastic systems. Nonetheless, the fact that
non-trivial exact results can be derived from the BE for IMP
justifies its study in order to gain a broader perspective on the
peculiar properties of dissipative gases. {}From that point of view,
the model \eqref{3.1} proposed here can be useful to have access, at
least at a semi-quantitative way, to relevant information (such as
the velocity distribution function itself) not directly available
from the BE for IMP.

The key ingredient of the kinetic model \eqref{3.1}, also present in
the BMD model \cite{BMD96}, is the  effective temperature
$\theta(\alpha) T<T$ in the reference distribution $f_0$. Its
existence leads to a non-trivial HCS solution with a positive
definite kurtosis and an algebraic high-energy tail. The additional
presence of the friction term relieves the effective temperature of
fully accounting for the cooling rate, so that a closer contact with
the BE for IMP is possible.

The analysis presented in this paper can be extended along a number
of routes. The explicit solution of the kinetic model to other
problems, such as the simple shear flow, is straightforward.
Moreover, the flexibility of the model allows one to choose its
parameters by a fit to other quantities, such as the effective
collision frequency $\nu_\kappa(\alpha)$ and/or the exponent
$s(\alpha)$, different from the ones considered in this paper. The
Gaussian kinetic model proposed in Ref.\ \cite{DBZ04}, which is able
to reproduce $\nu_\eta(\alpha)$ and $\nu_\kappa(\alpha)$
simultaneously, can be extended by the inclusion of a friction term,
thus providing an extra parameter $\gamma(\alpha)$. It s also
possible to generalize the model \eqref{3.1} to mixtures of IMP
\cite{GA05}.

%%%%%%%%%%%%%%%%%%%%%%%%%%%%%%%%%%%%%%%%%%%%%%%%
%% BACKMATTER
%%%%%%%%%%%%%%%%%%%%%%%%%%%%%%%%%%%%%%%%%%%%%%%%

\begin{theacknowledgments}
Partial support from the Ministerio de Educaci\'on y Ciencia
 (Spain) through Grant No.\ FIS2004-01399 is gratefully acknowledged.
\end{theacknowledgments}


%% If the bibliography is produced without BibTeX comment out the

\bibliographystyle{aipproc}   % if natbib is available

\end{document}